\date{Received: date / Revised version: date}

\documentclass[epj]{svjour}
\usepackage{epsfig,graphics,colordvi}
\usepackage{amsmath}
\usepackage{amssymb}
\usepackage{slashed}

\begin{document}

\title{Two-nucleon scattering:  merging chiral  effective field
  theory  with dispersion relations}
\author{A.M. Gasparyan\inst{1,2} \and M.F.M. Lutz\inst{3} \and E. Epelbaum\inst{1}
}  

\institute{Institut f\" ur Theoretische Physik II, Fakult\" at f\"ur Physik und Astronomie,
 Ruhr-Universit\" at Bochum 44780 Bochum, Germany 
\and SSC RF ITEP, Bolshaya Cheremushkinskaya 25, 117218 Moscow, Russia
\and GSI Helmholtzzentrum f\"ur Schwerionenforschung GmbH, Planckstra\ss e 1, 64291 Darmstadt, Germany}

\abstract{
We consider two-nucleon scattering close to threshold. Partial-wave amplitudes are obtained
by an analytic extrapolation of subthreshold reaction amplitudes
calculated in a relativistic formulation of chiral perturbation
theory. The constraints set by unitarity are used in order to stabilize the
extrapolation. Neutron-proton phase shifts are analyzed up to
laboratory energies $T_{{\rm lab}}\simeq250$ MeV based on the 
next-to-next-to-next-to-leading order expression for the subthreshold amplitudes.
We find a reasonably accurate description of the empirical S- and
P-waves and a good convergence of our approach. 
These results support the assumption that the subthreshold
nucleon-nucleon scattering amplitude may be computed perturbatively by
means of the chiral expansion. The intricate soft scales that govern the
low-energy nucleon-nucleon scattering are generated dynamically via a
controlled analytic continuation.    
}
\maketitle
\section{Introduction}

The last decade has witnessed an impressive progress towards a
quantitative solution of the nuclear many-body problem starting from
the underlying forces between the nucleons. Rapidly increasing
available computational resources coupled with modern ab-initio few-
and many-body methods as well as renormalization techniques to reduce 
the many-body model space make it nowadays possible, to carry out
reliable and accurate nuclear structure calculations for light and even 
medium-mass nuclei. One can therefore directly relate the fine
properties of the nuclear Hamiltonian to the spectra and other
properties of nuclei without invoking any uncontrollable
approximations. It is thus of utmost importance for contemporary
nuclear physics and nuclear astrophysics, to develop a
detailed, quantitative understanding of low-energy interactions
between the nucleons based on QCD, the underlying theory of the strong
force. Remarkable progress along these lines has been achieved in the
past two decades within the framework of (chiral) effective field theory. In
particular, accurate nucleon-nucleon (NN) potentials at
next-to-next-to-next-to-leading order in the chiral expansion have
been constructed \cite{Weinberg:1991um,Ordonez:1992xp,Entem:2003ft,Epelbaum:2004fk}. The corresponding developments for the three-nucleon
force are underway, see Refs.~\cite{Epelbaum:2008ga,Machleidt:2011zz} and references therein.  

As a complementary approach, NN scattering can also be  addressed from the
standpoint of the dispersion relations. This method has been
formulated and extensively explored  in the
sixties of the last century starting from the pioneering work by
Goldberger et al.~\cite{Goldberger:1960md}, see also
Refs.~\cite{Ball:1965sa,Scotti:1963zz,Scotti:1965zz}. Clearly, these 
studies were lacking a systematic theoretical framework to treat  
low-energy pion dynamics which is nowadays available in terms of 
chiral perturbation theory (ChPT). We also mention recent work
Refs.~\cite{Albaladejo:2011bu,Albaladejo:2012sa} along these lines,
where the impact of the left-hand cut emerging from 
one-pion exchange on NN scattering is investigated.  For recent
applications of dispersion relations in combination with chiral
perturbation theory to pion- and photon-nucleon dynamics the reader is referred to
Refs.~\cite{Gasparyan:2010xz,Gasparyan:2010fb,Gasparyan:2011yw,Ditsche:2012fv}.  

In the present paper, we use these ideas as developed in the recent works
\cite{Gasparyan:2010xz,Gasparyan:2010fb,Gasparyan:2011yw,Danilkin:2011fz,Danilkin:2012ua},
which we call dispersive effective field theory (DEFT),
to propose a novel approach to NN dynamics. It relies on the knowledge of the
analytic structure of the NN $S$-matrix and makes use of the
chiral expansion for the scattering amplitude, which is assumed to be valid in some
region below threshold, coupled with the conformal mapping techniques
to perform reliable extrapolations to higher energies. 

The dispersive approach formulated and applied in this
paper is complementary to methods based on  potentials derived 
within ChPT which have been extensively explored in the past decade. 
We expect it to bring new insights into various aspects related to
low-energy nucleon-nucleon scattering. First, given considerable
progress in the derivation of nuclear forces in the framework of
ChPT in recent years, it is important to unabiguously identify
effects of the long-range chiral two-pion exchange potential in 
nucleon-nucleon scattering data, see e.g. 
Refs.~\cite{Rentmeester:1999vw,Birse:2003nz,Rentmeester:2003mf,Birse:2007sx,Shukla:2008sp,Birse:2010jr}
for some studies along these lines based on the potential framework. 
The dispersive approach relies on an 
explicit evaluation of   the discontinuity across the low-lying
left-hand cuts in the amplitude and thus provides a transparent and 
efficient way to identify long-range physics. Secondly, there is 
an ongoing discussion in the community on how to carry out 
nonperturbative renormalization of the Schr\"odinger equation  in the
framework of chiral effective field theory and possible implications for power
counting, see \cite{Kaplan:1996xu,Lepage:1997cs,Beane:2001bc,Nogga:2005hy,Birse:2005um,PavonValderrama:2005wv,Epelbaum:2006pt,Long:2011xw}
for samples of different views. Our dispersive approach is based on
the on-shell scattering amplitude which, in the subthreshold region,
is constructed using chiral perturbation theory without performing any
nonperturbative resummations. Nonperturbative
effects in the physical region are generated by solving the integral
equation dictated by elastic unitarity.  
Our approach, therefore, allows to address the above-mentioned
issues from a completely different perspective and might shed new
light on this important problem. Last but not least, we also plan to
apply this method in the future to investigate the possibility to
treat the two-pion exchange potentials in perturbation theory
which is considered as an option to formulate renormalizable 
approaches to NN scattering in chiral EFT. 

Our manuscript is organized as follows. In section \ref{sec:method},
we briefly describe the theoretical method we are using to compute 
the NN scattering amplitude. The chiral expansion for the
discontinuities across the left-hand cuts is discussed in section
\ref{section3}. The results for NN phase-shifts are presented in
section \ref{sec:results}. Finally, our conclusions are summarized in
section \ref{sec:summary}. 

\section{An analytic continuation of the NN scattering amplitude}
\label{sec:method} 

\begin{figure}[tb]
\begin{center}
\includegraphics[width=8.8cm,keepaspectratio,angle=0,clip]{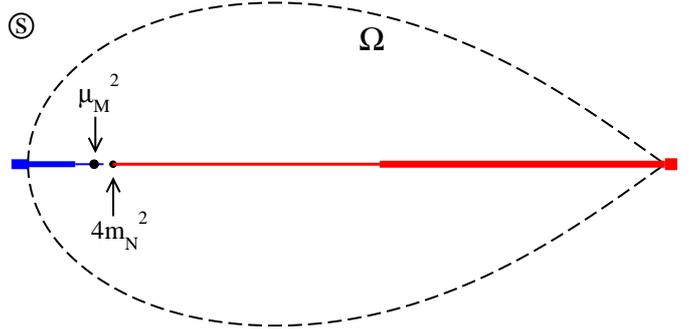}
\end{center}
\caption{Cuts of the partial wave NN amplitude in the complex
  $s$-plane.
The first, second and third left-hand cuts start at 
$s= 4m_N^2 -(n M_\pi )^2$, $n=1, 2, 3$.    
The first and second right-hand cuts due
to inelastic channels start at the one- and two-pion production
thresholds 
$s= (2 m_N +  n M_\pi )^2$ with $n=1, 2$. 
Further, $\mu_M$ denotes the
matching point as explained in the text.  
\label{fig:NN_cuts}}
\end{figure}

A NN partial-wave amplitude has a well-known singularity
structure as a function of the complex variable $s$. On the physical sheet, apart
from the pole corresponding to the deuteron bound state in the
$^3S_1-^3D_1$ channel, there are right- and left-hand 
cuts as shown in Fig.~\ref{fig:NN_cuts}. The right-hand cuts
correspond to intermediate states  
in the $s$-channel and start from the two-nucleon threshold,  $s=4m_N^2$.

The discontinuity of the amplitude across this cut is determined by the unitarity condition
\begin{eqnarray}
\Delta T(s)&\equiv&\frac{1}{2\,i}\left(T(s+i\epsilon)-
T(s-i\epsilon)\right)\nonumber\\
&=&T(s+i\epsilon)\,\rho(s)\,T(s-i\epsilon)+\dots\,,
\label{discontinuity}
\end{eqnarray}
where the dots stand for inelastic channel contributions which start at the pion production 
threshold,  $s=(2m_N+M_\pi)^2$. The phase-space function $\rho(s)$ and
the amplitude $T(s)$ in Eq.~(\ref{discontinuity})  
turn into $2 \times 2$ matrices for the coupled partial waves.

It is 
advantageous
to discriminate between the two sources of branch cuts in the
scattering amplitudes. The so-called generalized potential $U(s)$  
collects all contributions emerging from the left-hand cuts
only. Given the generalized potential, the full  
scattering amplitudes may be reconstructed in terms of the non-linear
integral equation 
\begin{equation}
 T(s\,)=U(s\,)+\, \int_{4m_N^2}^{\infty}\frac{ds'}{\pi}\,\frac{s-\mu_M^2}{s'-\mu_M^2}\,
\frac{T(s')\,\rho(s')\,T^*(s')}{s'-s-i\epsilon}\,.
\label{def-non-linear}
\end{equation}
Obviously, the solution of this equation  recovers the right-hand cut
in agreement with the elastic unitarity constraints. 
We further emphasize that the employed dispersion relations are
sometimes written in the literature using non-relativistic variables,
which is of course equivalent to the expressions used here.
Since we neglect the contributions from inelastic channels in
Eq.~(\ref{discontinuity}), $U(s)$ is well-defined only  
up to some $s=\Lambda_s^2$ below which the inelastic contributions are
small. We choose this value to correspond to the two-pion production threshold 
$\Lambda_s=2m_N+2M_\pi$. The dependence of our results on a
particular choice of $\Lambda_s$ will turn out to be rather weak, see
section \ref{sec:results}.  
At the matching point $s=\mu_M^2$,  the scattering amplitude $T(s)$
per construction equals the generalized potential $U(s)$. We choose
the matching point to be in the middle of the region between the
unitarity $s$-channel cut and two-pion $t$-channel cut,
i.e. $\mu_M^2=4m_N^2-2M_\pi^2$. The proper choice of the matching
point is essential for the effectiveness of our approach.  Our central
assumption is a perturbative nature of the scattering amplitude near 
the matching point. Consequently,  the generalized potential can be
calculated at the matching point in the framework of ChPT. Upon
performing an analytic continuation 
in the $s$-variable  by employing a suitable chosen conformal mapping technique, 
the generalized potential $U(s)$ can be used in Eq.~(\ref{def-non-linear}).

Consider the left-hand singularity structure of the generalized potential. The physically relevant cuts 
correspond to the $t$-channel one-pion and several-pion exchanges as
visualized in Fig.~\ref{fig:NN_cuts}.  These cuts start at $s=4m_N^2-(n M_\pi)^2$, where 
$n$ is the number of exchanged pions. We follow here the strategy of
\cite{Gasparyan:2010xz} and split the generalized potential into two parts,
\begin{equation}
U(s)=U_{\rm inside}(s)+U_{\rm outside}(s),
\end{equation}
where $U_{\rm inside}(s)$ is calculated explicitly whereas $U_{\rm
  outside}(s)$ is analytically continued. In Fig.~\ref{fig:NN_cuts}
we introduce a  
domain $\Omega$, which defines the separation of the 'inside' and
'outside' contributions to the generalized potential.  
The left and right-most point of the domain 
$\Omega$ are specified by $\Lambda^2$ (defined below) and $\Lambda_s^2$.
The discontinuity across the left-hand cut from $s=4m_N^2-M_\pi^2$
to  $s=4m_N^2-4 M_\pi^2$ is unambiguously determined by the
one-pion exchange contribution in terms of 
the $\pi NN$ coupling constant. It lies inside the domain $\Omega$ as
shown in Fig.~\ref{fig:NN_cuts}. 
The next part of the  cut from $s=4m_N^2-4M_\pi^2$ to
$s=4m_N^2-9 M_\pi^2$ receives, in addition, the contributions  
from the two-pion exchange processes. We assume that in this energy
region such contributions 
can be computed in ChPT and, therefore, they are also considered as
part of  $U_{\rm inside}(s)$. For $s< 4m_N^2-9 M_\pi^2$,  the two-pion-exchange 
cut can no longer be reliably computed in perturbation theory. This
reflects the nature of the $\pi \pi$ interaction in this kinematical
region with its  
prominent and non-perturbative scalar-isoscalar channel. Therefore,
we do not include this part into  $U_{\rm inside}(s)$ 
and identify  
\begin{eqnarray}
U_{{\rm inside}}(s)  &=&  \int_{\Lambda^2}^{4m_{N}^{2}-M_{\pi}^{2}}\frac{ds'}{\pi}\frac{\Delta T(s')}{s'-s}\,, \nonumber\\
 \Lambda^2&=&4m_N^2-9 M_\pi^2\,,
\end{eqnarray}
where $\Delta T(s)$ is the discontinuity of the amplitude across the
cut. By definition, the 'outside' part of the potential, $U_{\rm
  outside}(s)$,  
has singularities only outside the region $\Omega$ of
Fig.~\ref{fig:NN_cuts}. It incorporates, in particular,
effects of three  
pion exchanges that are technically more challenging to compute. Since
$U_{\rm outside}(s)$ is analytic inside  
$\Omega$, one may approximate it by a Taylor expansion around the NN
threshold. The expansion coefficients would resemble the infinite
tower of counter  
terms appearing in 
the chiral Lagrangian. However,  such a series would
converge only inside a circle with the radius $R=9M_\pi^2$ and,
therefore, 
the generalized potential will be not suitable for using as input in Eq.~(\ref{def-non-linear})
in order to compute the partial-wave scattering amplitudes in a controlled manner. 
We recall that the generalized potential is needed at least up to
$s\approx 4m_N^2+8m_N  M_\pi$ which  
is roughly $6$ times farther away from the NN threshold than
accessible by that Taylor expansion with convergence radius
$R=9M_\pi^2$.

Following \cite{Gasparyan:2010xz,Danilkin:2010xd,Danilkin:2011fz}, we
approximate $U_{\rm outside}(s)$ in the domain $\Omega$ of
Fig.~\ref{fig:NN_cuts} in a systematic way by means of an appropriate  
conformal transformation. A function $\xi(s)$ having inverse $\zeta(\xi)$ that maps $\Omega$ onto a unit circle with 
$\xi(4m_N^2)=0$ is constructed. The particular form of the conformal mapping is irrelevant. 
The potential $U_{\rm outside}(s)$ can now be expanded in a Taylor series in the conformal variable,
\begin{eqnarray}
U_{{\rm outside}}(s\,)&=&\sum\limits _{k=0}^{n}{U_{k}}\,\big[\xi (s\,)\big]^{k}\,,\nonumber\\
U_{k}&=&\frac{d^{k}U_{{\rm outside}}(\zeta(\xi))}{k!\, d\xi^{k}}\Big|_{\xi=0}\,,
\label{U_outside}
\end{eqnarray}
that can be proven to converge in the full domain $\Omega$ of
Fig.~\ref{fig:NN_cuts}.  
The coefficients $U_k$ are determined by derivatives of $U_{{\rm
    outside}}$ at the expansion point $s= 4m_N^2$.  
If the expansion in (\ref{U_outside}) is truncated at order $n$, the
first $n$-derivatives need to be computed.  
We choose the number of terms $n+1$ in Eq.~(\ref{U_outside}) to be equal to
the number of local counter terms contributing to the  
considered partial wave in ChPT at a given order so that there is a 
one-to-one correspondence between Eq.~(\ref{U_outside})  and the
chiral expansion.  

To be specific, we use here the conformal mapping suggested in
\cite{Gasparyan:2010xz}
\begin{eqnarray}
\xi(s)&=&\frac{a_1\left(\Lambda^2_s-s\,\right)^2-1}
{(a_1-2\,a_2)\left(\Lambda^2_s-s\,\right)^2+1}\,, 
\; \; \nonumber\\a_1&=&\frac{1}{\left(\Lambda^2_s-4m_N^2\right)^2}\,, 
\; \; 
a_2=\frac{1}{\left(\Lambda^2_s-\Lambda^2\right)^2}\,.
\label{mapping2}
\end{eqnarray}
A useful property of the transformation (\ref{mapping2}) is that it permits a smooth 
extrapolation of the generalized potential to a constant
above $s=\Lambda^2_s$. Replacing in (\ref{U_outside}) the conformal
map $\xi(s)$ by $\xi_>(s)$ defined as 
\begin{equation}
\xi_>(s\,)=\left\{
\begin{array}{ll}
\xi(s\,) \qquad & {\rm for} \qquad s<\Lambda^2_s\,, \\
\xi(\Lambda^2_s) \qquad & {\rm for} \qquad s\geq\Lambda^2_s \,,
\end{array} \right.  \quad \xi'(\Lambda^2_s) =0 \,,
\label{xiLambda}
\end{equation}
the outside potential is smoothly extended to a constant at energies $s> \Lambda_s^2$. 

The final step in obtaining the amplitude in the physical region is to
solve the integral equation~(\ref{def-non-linear}) with 
respect to $T(s)$ for a given $U(s)$. A standard way of doing that 
is the so-called $N/D$ technique \cite{Chew:1960iv}. 
The amplitude is represented as
\begin{equation}
T(s)=D^{-1}(s)N(s)\,,
\label{def-NoverD}
\end{equation}
where $D(s)$ has no singularities  but the right-hand $s$-channel unitarity cuts.
In contrast, the branch points of $N(s\,)$ correspond to those of $U(s\,)$.
The unitarity condition implies
\begin{equation}
D(s\,)=1-\int_{4m_N^2}^\infty \frac{d s'}{\pi}\,\frac{s-\mu_M^2}{s'-\mu_M^2}\frac{N(s')\,\rho(s')}{s'-s}\,.
\label{def-D}
\end{equation}
The non-linear equation~(\ref{def-non-linear}) reduces to the linear
one for $N(s)$, 
\begin{eqnarray}
N(s\,)&=&U(s\,)
 +\int_{4m_N^2}^\infty \frac{d s'}{\pi}\,
\frac{s-\mu_M^2}{s'-\mu_M^2}\,\nonumber\\ &\times&\frac{N(s')\,\rho(s')\,[U(s')-U(s\,)]}{s'-s}\,,
\label{Nequation0}
\end{eqnarray}
which can be solved using standard numerical techniques. 

It should be emphasized that apart from the dynamical singularities in
the complex $s$-plane discussed above,  the partial-wave amplitudes may have
kinematic singularities or obey kinematic constraints. In particular,
the standard $JLS$ amplitudes possess  $\sqrt{s}$-type singularities and
obey constraints at threshold, i.e. at $\sqrt{s}=\pm 2m_N$. Although,
apart from constraints at  
$\sqrt{s}=2m_N$, they are located rather far away from the physical region,
it is preferable  to remove all of them if possible. A general method
to achieve this goal for a system of two  
interacting spin-$1/2$ particles is presented in \cite{Stoica:2011cy} and some particular cases were discussed
in \cite{Ball:1965sa}.
We choose a set of amplitudes (which is generally speaking not unique) for the NN system as follows.
For the uncoupled partial waves with $J=L$ and the $^3P_0$ channel,
the $S$-matrix elements $S^{JLS}(s)$ or, equivalently,  the phase
shifts $\delta^{JLS}(s)$ are related with the 
amplitude $T^{JLS}(s)$ which is free of the kinematical singularities
via  
\begin{equation}
 S^{JLS}(s)=e^{2\,i\,\delta^{JLS}(s)}=1+2\,i\,\rho_L(s)\,T^{JLS}(s)\,,
\end{equation}
where the phase-space function is defined according to 
\begin{equation}
\rho^{L}(s)=
\frac{1}{8\pi}\left(\frac{1}{4}-\frac{m_N^2}{s}\right)^{L+\frac{1}{2}} \,.
\end{equation}
For coupled partial waves, one needs, in addition, to perform 
a linear transformation of the $L=J\pm 1$ basis leading to the
following relation in terms of the Stapp parametrization
\cite{Stapp:1956mz}
\begin{eqnarray}
 S^J&=&\left(\begin{matrix}
\cos{2\epsilon_J}e^{2i\delta^{J-1}} &i\sin{2\epsilon_J}e^{i(\delta^{J-1}+\delta^{J+1})}  \\
i\sin{2\epsilon_J}e^{i(\delta^{J-1}+\delta^{J+1})} &\cos{2\epsilon_J}e^{2i\delta^{J+1}}
 \end{matrix}\right) \nonumber\\[4pt]
&=&1+2\,i\,\frac{1}{8\pi}\left(\frac{1}{4}-\frac{m_N^2}{s}\right)^{J-\frac{1}{2}}\,
U^T \,T^{J}  \, U\,,
\end{eqnarray}
with the transformation matrix
\begin{equation}
U=\left(\begin{matrix}
\frac{2 (J+1) m_N}{J \sqrt{s}}+1 &\sqrt{\frac{J+1}{J}}
 \left(1-\frac{2m_N}{\sqrt{s}}\right)\\
 \sqrt{\frac{J+1}{J}} \left(\frac{1}{4}-\frac{m_N^2}{s}\right)&
   \frac{m_N^2}{s}-\frac{1}{4}\,,
 \end{matrix}\right)\,,
\end{equation}
which implies that the NN phase-space distribution has
the form
\begin{eqnarray}
\rho^{J}&=&\frac{2\,J+1}{8\pi J}\left(\frac{1}{4}-\frac{m_N^2}{s}\right)^{J-\frac{1}{2}}\nonumber\\
&\times&
\left(\begin{matrix}
 1+\frac{4m_N^2}{s}\frac{J+1}{J}& 2\,\sqrt{\frac{J+1}{J}}\left(\frac{1}{4}-\frac{m_N^2}{s} \right) \frac{m_N}{\sqrt{s}} \\
2\,\sqrt{\frac{J+1}{J}}\left(\frac{1}{4}-\frac{m_N^2}{s} \right) \frac{m_N}{\sqrt{s}}&
   \left(\frac{1}{4}-\frac{m_N^2}{s}\right)^2
 \end{matrix}\right).\nonumber
\end{eqnarray}

\section{Chiral perturbation theory for the left-hand cuts}
\label{section3}

We now discuss the derivation of the generalized potential in
ChPT up to order $Q^3$ with $Q\sim M_\pi$
referring to a generic soft scale. 
The following terms in the effective Lagrangian are relevant  for our
calculation: 
\begin{eqnarray}
\mathcal{L}_{int} & = & -\frac{1}{4\, F_\pi^{2}}\,\bar{N}\,\gamma^{\mu}\,\big(\vec{\tau}\cdot\big(\vec{\pi}\times(\partial_{\mu}\vec{\pi})\big)\big)\, N\nonumber\\
&+&\frac{g_{A}}{2\, F_\pi}\,\bar{N}\,\gamma_{5}\,\gamma^{\mu}\,\big(\vec{\tau}\cdot(\partial_{\mu}\vec{\pi})\big)\, N\nonumber\\
&-&\frac{2\,c_1}{F_\pi^2}\,M_\pi^2\, \bar{N}\,( \vec{\pi}\cdot\vec{\pi})\,N
+ \frac{c_3}{F_\pi^2}\,\bar{N} \,(\partial_{\mu}\,\vec{\pi} )
\cdot (\partial^{\mu}\vec{\pi})\,N
\nonumber \\
&-& \frac{c_4}{2\,F_\pi^2}\,\bar{N}\,\sigma^{\mu\nu}\,\big(\vec{\tau} \cdot \big((\partial_{\mu}\vec{\pi})\times
(\partial_{\nu}\vec{\pi})\big)\big)\,N\,.
\label{def-Lagrangian}
\end{eqnarray}
Here, $\vec  \pi$ and $N$ refer to pion and nucleon fields, $\vec
\tau$ denote the isospin Pauli matrices while $F_\pi$ and $g_A$ are the
pion decay and the nucleon axial vector constants,
respectively. Further, $c_i$ are the low-energy constants (LECs) accompanying
the subleading pion-nucleon vertices. For more details on the
effective pion-nucleon Lagrangian the reader is referred to
Refs.~\cite{Fettes:1998ud,Fettes:2000gb}.  

\begin{figure}[tb]
\begin{eqnarray*}
T^{(0)}&=&
\parbox{1.8cm}{\includegraphics[width=1.8cm,clip=true]{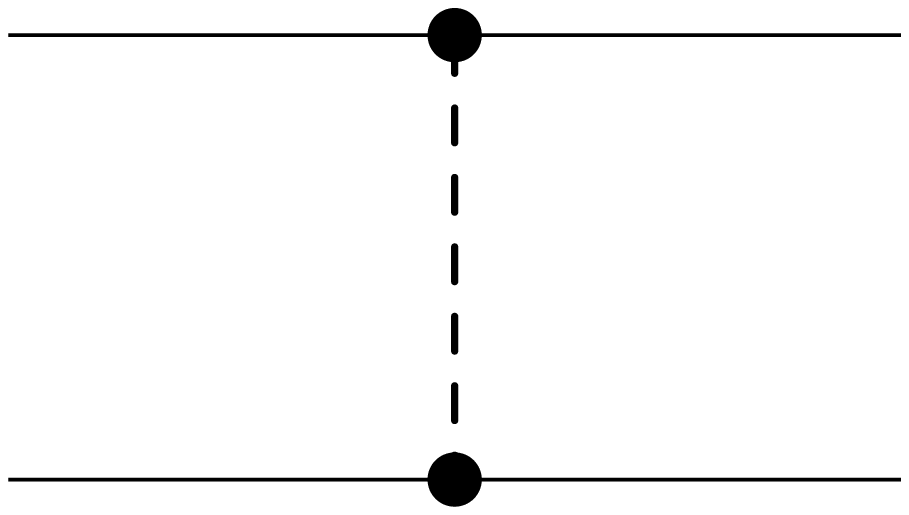}}+\dots\,,  \qquad 
T^{(1)}=\parbox{1.8cm}{\includegraphics[width=1.8cm,clip=true]{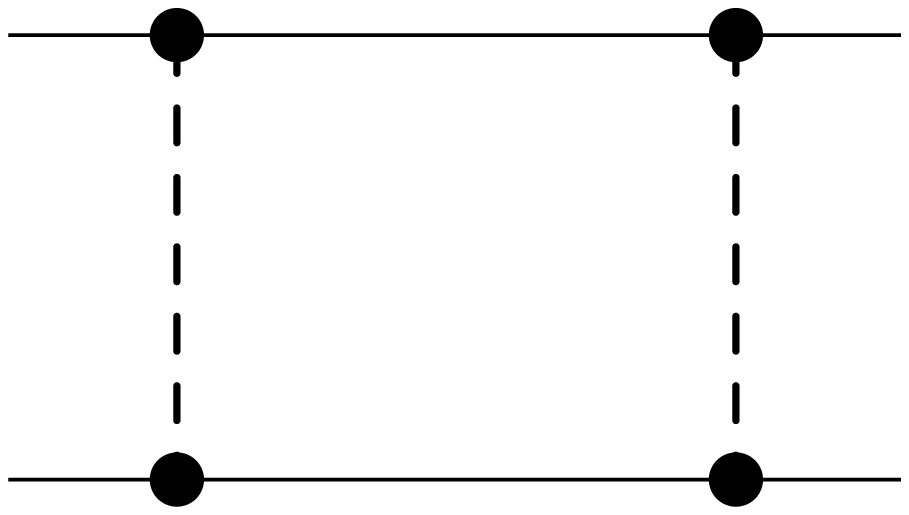}}+\dots\,,\\[10pt]
T^{(2)}&=&\parbox{1.8cm}{\includegraphics[width=1.8cm,clip=true]{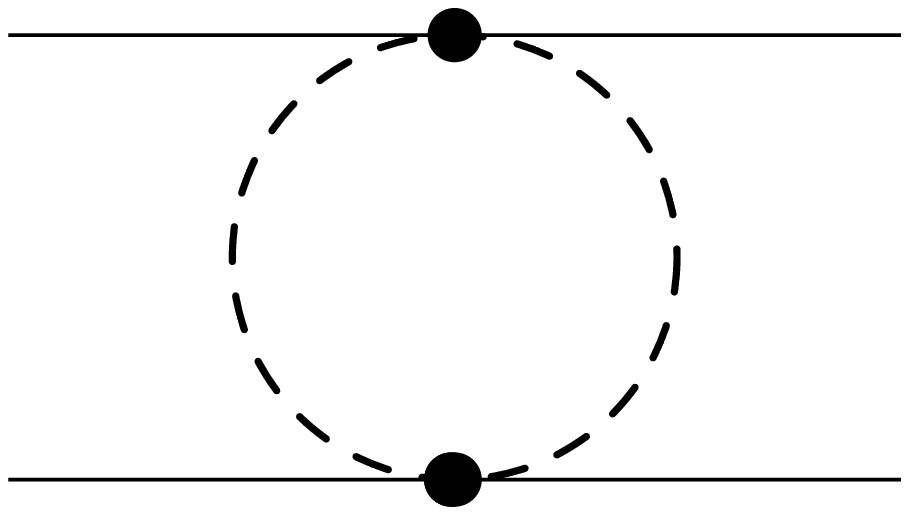}}
+\parbox{1.8cm}{\includegraphics[width=1.8cm,clip=true]{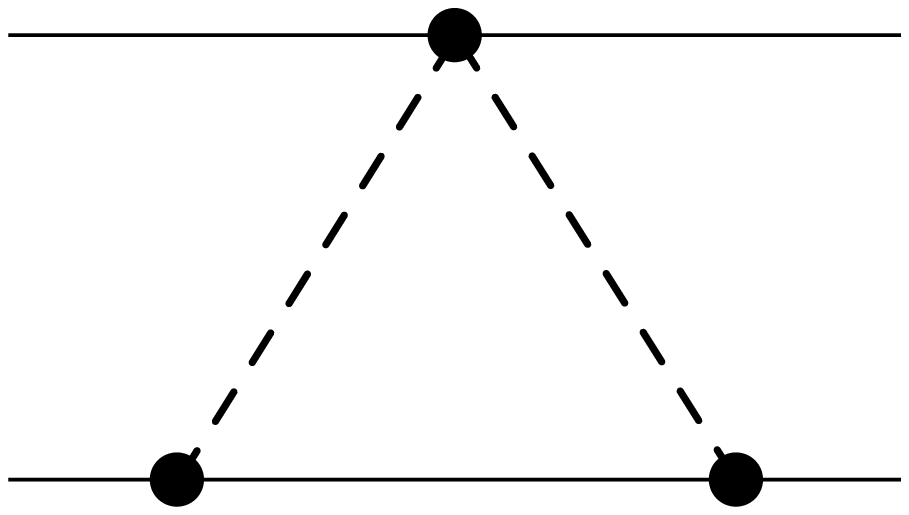}}
+\parbox{1.8cm}{\includegraphics[width=1.8cm,clip=true]{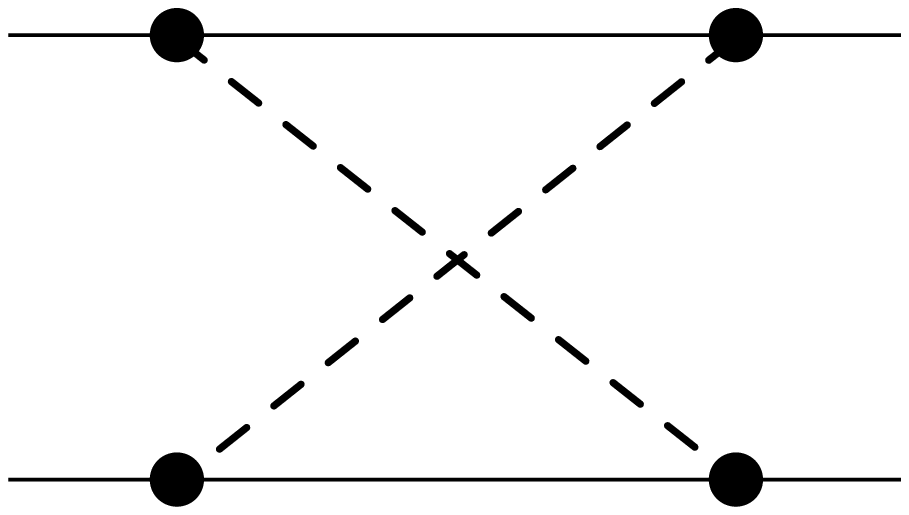}}+\dots\,,\\[10pt]
T^{(3)}&=&\parbox{1.8cm}{\includegraphics[width=1.8cm,clip=true]{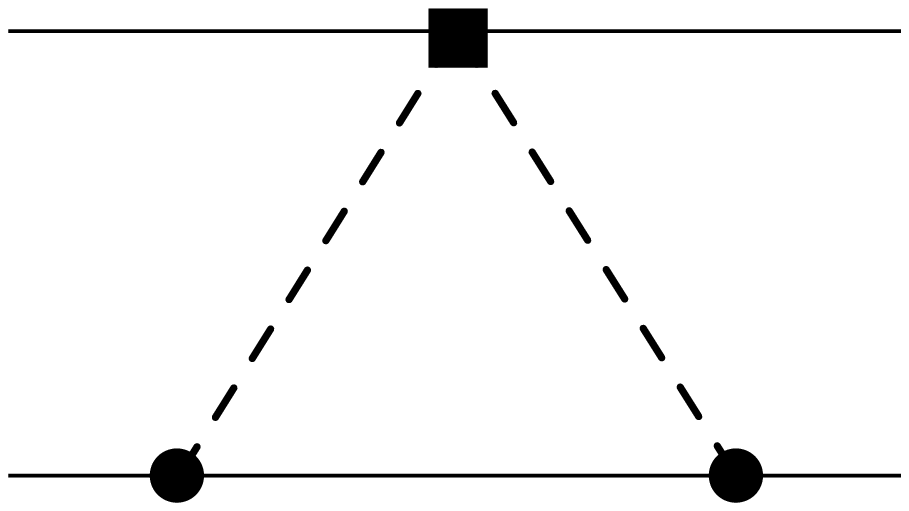}}+\dots
\end{eqnarray*}
\caption{Feynman diagrams contributing to the NN scattering
  amplitude $T^{(i)}$ at order $Q^i$ in the chiral expansion.
Solid dots (filled squares) denote the lowest-order vertices in the
first line of Eq.~(\ref{def-Lagrangian}) (subleading vertices $\propto
c_i$). The ellipses denote diagrams leading to shorter-range
contributions.  
\label{fig:diagrams}}
\end{figure}

We apply  the effective Lagrangian introduced above to compute the NN
scattering amplitude to order $Q^3$. As already pointed out before,
our central assumption is perturbativeness of the scattering amplitude
in the region $s \sim \mu _M^2$ (see Ref.~\cite{Danilkin:2010xd} for a quantum-mechanical example).  We, therefore, use here the standard
chiral power counting based entirely on the naive dimensional analysis.
\footnote{Notice that nonperturbative renormalization of the Schr\"odinger equation and implications for the 
chiral EFT power counting are currently under discussion, see \cite{Kaplan:1996xu,Lepage:1997cs,Beane:2001bc,Nogga:2005hy,Birse:2005um,PavonValderrama:2005wv,Epelbaum:2006pt,Long:2011xw}
for samples of different views.} 
Although we employ the manifestly Lorentz-invariant form of
$\mathcal{L}_{int}$ without performing the $1/m_N$-expansion in order not to distort the  discontinuity structure across
the left-hand cuts, we apply the standard power counting rules of
the heavy-baryon formulation and keep only 
those diagrams which do not vanish in the heavy-baryon framework at a given order.
The Feynman diagrams emerging at various orders in the chiral
expansion are depicted in Fig.~\ref{fig:diagrams}.
Notice that the shorter-range contributions emerging from graphs which
are not
shown explicitly in Fig.~\ref{fig:diagrams} including loop diagrams with contact interactions,
$3$- and $4$-pion exchange diagrams do not generate  
left-hand cuts inside the $\Omega$-domain and,  therefore, can only influence the 
normalization of the amplitude at the matching point. For the S- and
P-waves, these effects are absorbed in the corresponding counter
terms which are available at order $Q^3$.   For higher partial waves
one, in principle, needs to evaluate such diagrams explicitly, at
least at the NN threshold. We remind the reader that at this order, 
the generalized potential in partial waves with $L+L'>2$, where $L$ ($L'$) denotes the initial
(final) orbital angular momentum, contains only the constant term $U_0$ in the $\xi$-expansion. 
The zeroth-order coefficient $U_0$ in the expansion of the "outside" part of the generalized potential 
for higher partial waves emerges at order $Q^3$ from three- and four-pion exchange diagrams (iterations 
of the one-pion exchange potentials in the conventional framework). The corresponding left-hand cuts 
are outside the $\Omega$-domain. Therefore, we expect these contributions to $U_0$ at the matching point 
($\mu_M=4m_N^2-2M_\pi^2$) to be suppressed relative to the long-range contributions by a factor of 
$((2M_\pi^2)/(9M_\pi^2))^L = (2/9)^L$ and neglect them in the present analysis. 

Given explicit expressions for the perturbative scattering amplitude,
the generalized potential can be obtained using the framework outlined
in the previous section. The explicit calculation of two-pion exchange
diagrams in the covariant framework is carried out e.g.~in
Refs.~\cite{Lutz:1999yr,Higa:2003jk}. For calculating 
discontinuities it is convenient to employ the Passarino-Veltman
reduction of the  
amplitude \cite{Passarino:1978jh,Lutz:1999yr,Semke:2005sn} in terms of
the scalar loop integrals, for which simple dispersion representations
are available, see e.g.~\cite{Lutz:1999yr}. 
Renormalization of the NN amplitude requires a special care in the
covariant framework. We follow here the strategy of Ref.~\cite{Fuchs:2003qc}. 
After applying dimensional regularization in the $\overline{MS}$  scheme,
an additional subtraction is done in order to restore the power counting.  
The subtracted terms can be absorbed by lower-order counter
terms. In practice, it is convenient to use the prescription proposed  
in \cite{Lutz:1999yr,Semke:2005sn} in which the subtraction is carried
out directly at the level of the scalar loop integrals. 
The details of this calculation will be published
separately.  Finally, we emphasize that $1/m_N$-expansion of the derived expressions for
the two-pion exchange amplitude leads to results consistent with the ones given in
Refs.~\cite{Kaiser:1997mw,Friar:1999sj}.

\section{Results for neutron-proton phase shifts}
\label{sec:results}

We are now in the position to discuss our numerical results for
neutron-proton phase shifts. In this work, we restrict ourselves to
low partial waves where non-perturbative effects are most pronounced.
In each channel, we solve the nonlinear integral equation
(\ref{def-non-linear}) numerically using the $N/D $ method as
  outlined in section \ref{sec:method}.  The $U_{\rm inside} (s)$-part
  of the generalized potential depends on the LECs $F_\pi$, $g_A$ and
  $c_i$. Here and in what follows, we adopt the value 
$F_\pi \simeq 92.4$ MeV  for the pion decay constant. Further, we
implicitly account for the Goldberger-Treiman discrepancy by using the
effective value for $g_A$, 
\begin{equation}
g_A = \frac{F_\pi \, g_{\pi NN}}{m_N} \simeq 1.285\,
\end{equation}
based on $g_{\pi NN}^2/(4 \pi) \simeq 13.54$
\cite{Timmermans:1990tz,Baru:2010xn}. 
Such a replacement is legitimate to the order we are working and allows
us to take 
into account 
the one-pion exchange discontinuity exactly. The
sensitivity of our fits to the precise value of $g_{\pi N N}$ will be
discussed below. By the same reason,  we always distinguish between
the neutral and charged pion masses in the one-pion exchange
contribution although formally these isospin-breaking effects start
at order $Q^2$. The corrections due to the different pion masses in the box diagram appear
at order $Q^3$. We include these corrections perturbatively following the
lines of Ref.~\cite{Friar:2003yv}. The impact on phase shifts is,
however, almost negligible. Last but not least, for the LECs $c_{1,3,4}$, we adopt the
following values from the fit of
Ref.~\cite{Krebs:2012yv}: $c_1=-0.75\,\rm{GeV}^{-1}$, $c_3=-4.77\,\rm{GeV}^{-1}$, $c_4=3.34\,\rm{GeV}^{-1}$.
Interestingly, we observe that effects of
the order-$Q^3$ contributions proportional  
to  $c_i$'s are fairly small for the partial waves considered. In
particular, varying these constants within the limited range
suggested in the literature does not significantly affect the quality
of the fit.  These findings are in line with the ones of 
Ref.~\cite{Shukla:2008sp} and are related to the fact
that we explicitly take into account only the long-range part of the two-pion exchange
contributing to $U_{\rm inside}$ and absorb the rest into the coefficients of the conformal-mapping
expansion. From the conceptual point of view, this is similar to 
the approach suggested in
Refs.~\cite{Epelbaum:2003gr,Epelbaum:2003xx} based on the 
spectral-function regularization of the two-pion-exchange potential. 
Using this framework, the contributions from the short-range part of
the two-pion-exchange 
are suppressed. 

Note, that constants $c_3$ and $c_4$ contain contributions from 
the $\Delta$ resonance \cite{Bernard:1996gq} which is  at present  not included in
our Lagrangian as an explicit degree of freedom.

Finally, the coefficients $U_k$ entering the truncated
expansion of the $U_{\rm outside} (s)$-part of the generalized
potential are determined by fitting the 
empirical phase shifts of the Nijmegen partial wave analysis (PWA)
\cite{Stoks:1993tb} up to $T_{lab}=100$ MeV.  The behavior of phase
shifts at higher energies comes out as a prediction.   Notice that the 
number of free parameters in a given partial wave is determined by the number of local NN counter terms in the Lagrangian 
relevant at a given  order. At even orders $Q^n$ ($n=0,2$) there are counter terms contributing to  the partial waves with $L+L'=n$. 
The explicit form of the short-range NN terms in the Lorentz-invariant Lagrangian can be found in Ref.~\cite{Girlanda:2010zz}.
The parameters of the fit are specified in appendix~\ref{appendix}. Instead of giving explicitly the values of the 
coefficients of the $\xi$-expansion we provide the values of the generalized potential and its derivatives at the matching point
needed to unambiguously deduce these coefficients.
Those quantities are renormalization-scale independent.
We further subtract the large one-pion-exchange contribution, which is uniquely defined.

\begin{figure*}[tb]
\begin{center}
\includegraphics[width=14.0cm,keepaspectratio,angle=0,clip]{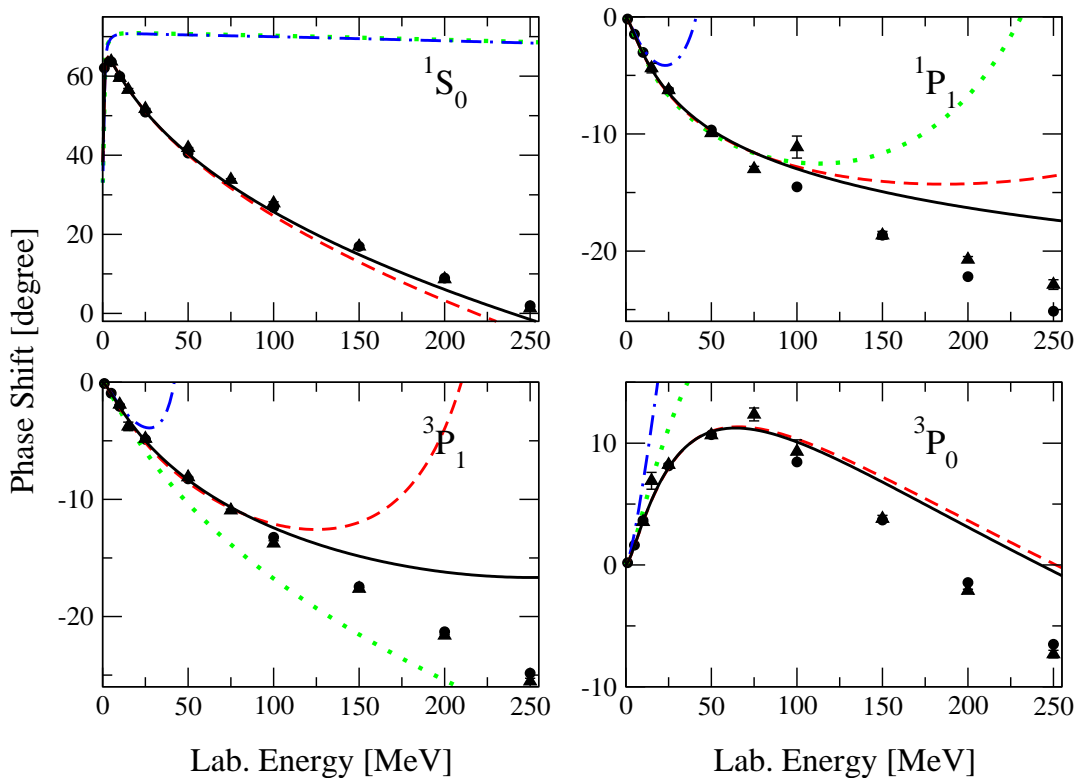}
\caption{Phase shifts in uncoupled S- and P-waves calculated at orders
  $Q^0$ (dotted lines), $Q^1$ (dash-dotted lines), $Q^2$ (dashed
  lines) and $Q^3$ (solid lines) in comparison with the  Nijmegen
  \cite{Stoks:1993tb} (filled circles) and SAID \cite{SAID_NN} (filled
  triangles) PWAs. Notice that the dotted and dash-dotted lines nearly
  coinside for the $^1$S$_0$ phase shift. 
 }
\label{fig:1S0_1P1_3P1_3P0}
\end{center}
\end{figure*}

Our results for uncoupled channels are shown in Fig.~\ref{fig:1S0_1P1_3P1_3P0}.
In the $^1S_0$ partial wave,  there is one free parameter (two free
parameters) at orders $Q^0$ and $Q^1$ ($Q^2$ and $Q^3$). 
It is well known that an accurate description of the $^1S_0$ phase
shift requires the inclusion of the subleading contact interaction in
order to account for a rather large effective range in this
channel. Our results agree with these expectation. In particular, we
observe a good description of the data at orders $Q^2$ and $Q^3$,
where two fit parameters are available.  Interestingly, including only 
the first left-hand cut  $4m_N^2-4M_\pi^2 < s <    
4m_N^2-M_\pi^2$ explicitly in $U_{\rm inside}$ leads to fits of a comparable
quality.  On the other hand, the fit appears to be rather sensitive to the value of
$g_{\pi NN}$ coupling constant. Treating it as a free parameter yields
a value for $g_{\pi NN}^2$ which lies within $10$ percent 
of the empirical one if one fits the data up to $T_{lab}=100$
MeV. This difference gets even smaller if the fit is restricted to a
narrower energy region around the threshold. A similar sensitivity to
the value of $g_{\pi NN}$ is observed in the uncoupled  P-waves of Fig.~\ref{fig:1S0_1P1_3P1_3P0}.

For P-waves, there are no free parameters (one free parameter) at
orders $Q^0$ and $Q^1$ ($Q^2$ and $Q^3$). The calculated phase shifts
at orders $Q^0$ and $Q^1$ ($Q^1$) in the $^1P_1$ ($^3P_1$)
partial wave show a steep rise already at rather low energies. This
reflects the appearance of nonphysical resonances  
generated by solving the $N/D$ equation. These artifacts disappear at
orders $Q^2$ and $Q^3$.  In the $^3P_0$ channel we observe a large
contribution of the lowest-order counter term even at low energies
which agrees qualitatively with the findings
of Refs.~\cite{Nogga:2005hy,Epelbaum:2012ua}.  
Given that there is just one free parameter in each of the P-wave, the
phase shifts are remarkably well reproduced at order $Q^3$.

\begin{figure*}[tb]
\begin{center}
\includegraphics[width=14.0cm,keepaspectratio,angle=0,clip]{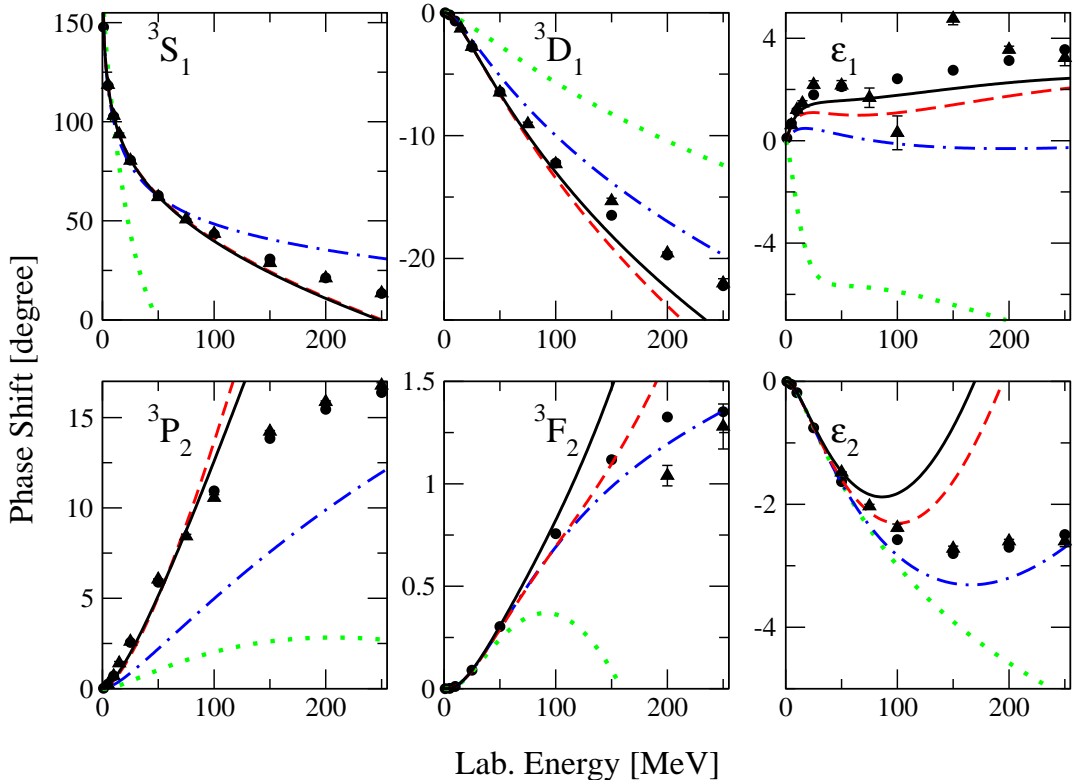}
\caption{Phase shifts and mixing angles in the coupled
  $^3S_1$-$^3D_1$ and   $^3P_2$-$^3F_2$ channels. 
For notation see Fig.~\ref{fig:1S0_1P1_3P1_3P0}.
 }
\label{fig:3S1_3D1_E1_3P2_3F2_E2}
\end{center}
\end{figure*}

We now turn to the coupled channels, see
Fig.~\ref{fig:3S1_3D1_E1_3P2_3F2_E2}, and first address the convergence
pattern in the $^3S_1$-$^3D_1$ partial waves. 
The lowest-order (i.e.~$Q^0$) results show a large disagreement with the
empirical phase shifts. In particular, the mixing angle $\epsilon_1$
even comes with the wrong sign. The order-$Q^1$ corrections 
emerging from the box diagram, see Fig.~\ref{fig:diagrams}, are large
and lead to a strongly improved description of the data. Notice that
there is just one free parameter at both orders $Q^0$ and $Q^1$. This 
suggests that the long-range physics associated with exchange
of pions plays a prominent role in this channel which agrees well with  
the results based on chiral and phenomenological NN potentials.  
Taking into account the $Q^2$- and $Q^3$-contributions leads to 
further improvement (in part, on the cost of having two more
adjustable parameters). 
The deuteron binding energy is also rapidly converging to its experimental
value $E_d=-2.22$ MeV.
In particular, we obtain $E_d=-32.70$ MeV, $-2.73$ MeV, $-2.10$ MeV  and $-2.12$ MeV at  orders
$Q^0$, $Q^1$, $Q^2$ and $Q^3$, respectively.
The situation in the $^3P_2$-$^3F_2$ is qualitatively similar: one
observes a large disagreement with the data at order $Q^0$ which is
strongly reduced by taking into account the corrections at order
$Q^1$.  Notice that the results at this order are parameter-free predictions.
The higher-order corrections bring our results into a better agreement
with the data (on the cost of adjusting a single  free 
parameter to the $^3P_2$ phase shift). The agreement with
the empirical phase shifts at order $Q^3$ is reasonably good,
especially given that there is only  one adjustable parameter at this order.  
Note that because of the channel coupling the fit is performed simultaneously for the three phases
that sometimes leads to  situations when although the overall fit improves the results
get worse for some particular phases when going to higher chiral order.

\begin{figure*}[tb]
\begin{center}
\includegraphics[width=14.0cm,keepaspectratio,angle=0,clip]{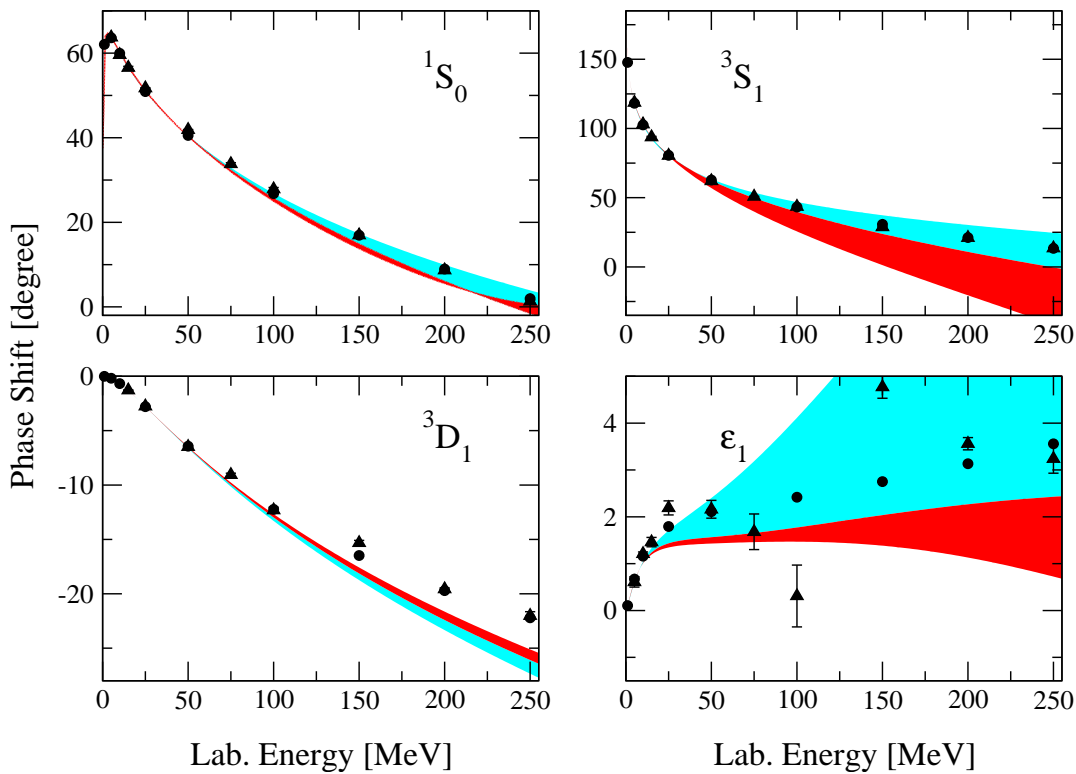}
\caption{Sensitivity of the $^1S_0$ and  $^3S_1$-$^3D_1$ partial waves
  to the variation of $\Lambda_s$ at order $Q^3$.  Light shaded bands correspond to the variation
of $\Lambda_s$ in the range $2m_N+M_\pi  \leq \Lambda_s \leq
2m_N+2M_\pi$, whereas dark shaded bands correspond to the range
$2m_N+2 M_\pi \leq
\Lambda_s < \infty$.}
\label{fig:1S0_3S1_3D1_E1Lambdas}
\end{center}
\end{figure*}

Apart from the counter terms, there are several parameters in our
scheme whose particular values were set by physical arguments. 
First, the matching scale $\mu_M$ should, of course, be taken far away
from the non-perturbative regions in the $s-$ and $t$-channels. That
is why we choose $\mu_M$ to lie between $s$- and $t$-channel two-body
thresholds. At sufficiently high order in the chiral  
expansion, the results should become independent on the particular
choice of $\mu_M$ since the dependence on $\mu_M$ can be compensated
by local NN counter terms. At order $Q^3$, we observe a rather week
dependence on $\mu_M$ in most of the partial waves we have studied.  
For example, varying the position of $\mu_M$ within the region $4m_N^2-3.5M_\pi^2<\mu_M^2<4m_N^2-1.1M_\pi^2$ affects the $^3D_1$
phase shifts at $E_{\rm lab} =250$~MeV, where the effect is the largest, by less then 10\%.
The variation in the leading order is, as one can expect, stronger but
does not exceed $30\%$ at $E_{\rm lab} \leq 100$~MeV. 
The second source of uncertainty is associated with the choice of 
conformal mapping and, in particular, of the value of $\Lambda_s$.
To the order we are working, the variation of $\Lambda_s$ only affects
the S-waves, where the expansion of $U_{\rm outside}$ goes 
beyond the zeroth order in $\xi$. Small variations of $\Lambda_s$ do
not lead to sizable effects on the results. 
For an illustration, we show in Fig.~\ref{fig:1S0_3S1_3D1_E1Lambdas}
the effect of the variation of $\Lambda_s$ within the extremely large
range $2m_N+M_\pi \leq \Lambda_s < \infty$. 
Both limits are highly unphysical:  the lower one is located very
close to the energy region we are interested in and thus makes the  
$\xi$-expansion there unreliable while the upper one assumes the
irrelevance of inelastic cuts in the whole energy region. 
One can see that even within these extreme limits, the variation of the
phase shifts is rather modest. Only in the $^3S_1$ phase shift and in 
the mixing angle $\epsilon_1$ one observes a sizable variation at
higher energies. Varying $\Lambda_s$  within a reasonable range 
does not affect our results in a significant way. The dependence of the results on the parameter 
of conformal mapping $\Lambda$ is much weaker than the dependence on $\Lambda_s$ and will
be given in a separate publication.

We now briefly compare our results with those obtained in Refs.~\cite{Albaladejo:2011bu,Albaladejo:2012sa}. 
Differently to these calculations, an essential ingredient of our approach is
an analytic continuation of the generalized potential from the
subthreshold to the physical region by means of conformal mapping
which relies on a clear separation between soft and hard left hand
cuts. This also allows us to avoid a sophisticated regularization
procedure for the nonlinear integral equation Eq.~(\ref{def-non-linear}) adopted in
Refs.~\cite{Albaladejo:2011bu,Albaladejo:2012sa}. It is difficult to carry out a more precise comparison between our
approach and the one of Refs. [9,10] as those calculations are limited
to LO. We, however, emphasize that the results for the uncoupled
channels at LO look similar in both approaches. 

\section{Summary}
\label{sec:summary}

In this work, we performed an analytic continuation of the
subthreshold NN amplitude into the physical region 
assuming the validity of ChPT for the amplitude at some
matching point below threshold. The amplitudes at the matching point
and the discontinuities across the nearby left-hand cuts are calculated
to order $Q^3$ using a manifestly covariant version of ChPT.  
The contributions from distant left-hand cuts are extrapolated to
higher energies using conformal mapping techniques. The resulting
generalized potential is used to reconstruct the full scattering
amplitude by means of partial-wave dispersion relations supplemented
by the unitarity  constraint. Within our approach, the short-range
part of the generalized potential is parametrized in a systematic way
in terms of the Taylor expansion of $U_{\rm outside} (s)$  in a
conformal variable $\xi (s)$. From the conceptual point of view, this
is analogous to the representation of the short-range part of the
nuclear force in terms of contact interactions with an increasing
number of derivatives. We have determined the corresponding 
coefficients entering the expansion of $U_{\rm outside} (s)$ in S- and
P-waves by fitting them to the Nijmegen PWA up to the
energy of $T_{lab}=100$ MeV. The obtained fits and predictions for phase shifts
at higher energies are in a reasonable agreement with the empirical PWAs
up to $T_{lab}=250$ MeV. The quality of the fit at order $Q^3$ is comparable with the
one obtained in conventional approaches (at next-to-next-to-leading order) based on the chiral expansion
of the NN potential. In all partial waves, the expansion for the phase
shifts seems to converge when going from the order $Q^0$ to
$Q^3$. This supports our assumption of the validity
of a perturbative expansion of the NN amplitude at the matching point. 
We also studied the dependence of the results on the particular choice
of the matching point $\mu_M$ and the conformal mapping parameter
$\Lambda_s$. We found our results to be rather weakly dependent on $\mu_M$ and $\Lambda_s$
provided these parameters are varied within a physically acceptable range. 

Our detailed analysis provides new insights into the relative
importance of contributions associated with the one- and two-pion
exchange cuts. Such conclusions are more difficult to draw in 
calculations based on a potential scheme, where the left-hand cuts in
the scattering amplitude emerge not only from the potential but also
from its iterations within the dynamical equation. As an important outcome of
our investigation, we found unambiguously that NN scattering in the
$^3S_1-^3D_1$ channel shows a strong evidence of the lowest-order
two-pion exchange which in the  conventional,
Schr\"odinger-equation-based  approaches corresponds to the first
iteration of the one-pion exchange potential. In particular, taking
into account the leading two-pion exchange is absolutely necessary to
achieve a 
good description of the mixing angle $\epsilon_1$.

\section*{Acknowledgments}

We are grateful to Jambul Gegelia, Hermann Krebs and Daniel Phillips for useful discussions. 
This work is supported by the EU HadronPhysics3 project ``Study of strongly interacting matter'', 
by the European Research Council (ERC-2010-StG 259218 NuclearEFT) and 
by the DFG (TR 16, ``Subnuclear Structure of Matter''). 

\appendix
\section{Values of the generalized potential at the matching point}
\label{appendix}
The values of the generalized potential at the matching point are collected in tables~\ref{table:parameters1},\ref{table:parameters2},
whereas derivatives of the relevant matrix elements of the generalized potential at the matching point
are equal to 
\begin{eqnarray}
 ^1S_0: \ \frac{dU(s)}{ds}{\bigg\rvert}_{s=\mu_M^2}&=&-4.57\times 10^{-3}\,\rm{MeV}^{-2}\,, \nonumber\\
^3S_1-^3D_1: \ \frac{dU_{11}(s)}{ds}{\bigg\rvert}_{s=\mu_M^2}&=&-1.39\times 10^{-3}\,\rm{MeV}^{-2}\,, \nonumber
\end{eqnarray}
at order $Q^2$,
and
\begin{eqnarray}
 ^1S_0: \ \frac{dU(s)}{ds}{\bigg\rvert}_{s=\mu_M^2}&=&-2.71\times 10^{-3}\,\rm{MeV}^{-2}\,, \nonumber\\
^3S_1-^3D_1: \ \frac{dU_{11}(s)}{ds}{\bigg\rvert}_{s=\mu_M^2}&=&-1.45\times 10^{-3}\,\rm{MeV}^{-2}\,, \nonumber
\end{eqnarray}
at order $Q^3$. One-pion-exchange contributions are always subtracted.

\begin{table}[tb]
\begin{center}
\begin{tabular}{|c|c|c|c|c|c|c|}
\hline 
 & $^1S_0$  & $^1P_1$  & $^3P_1$  & $^3P_0$  \tabularnewline
\hline
$Q^0$&$5.79\times 10^2  $&  &  &   \tabularnewline
\hline
$Q^1$&$5.82\times 10^2  $&  & &  \tabularnewline
\hline
$Q^2$&$8.53\times 10^2 $&$1.42\times 10^3  $  &$3.46\times 10^3$& $-6.43\times 10^3 $ \tabularnewline
\hline
$Q^3$&$8.63\times 10^2 $&$2.19 \times 10^3 $ &$4.81 \times 10^3$ &$-5.70\times 10^3$  \tabularnewline
\hline
\end{tabular}
\end{center}
\caption{Generalized potential at the matching point $U(\mu_M^2)$ that contain
adjusted parameters at different chiral orders after subtracting one-pion exchange contribution.
Uncoupled partial waves.}
\label{table:parameters1}
\end{table}

\begin{table}[tb]
\begin{center}
\begin{tabular}{|c|c|c|c|c|c|c|}
\hline 
 & $^3S_1-^3D_1\,, \, U_{11}$  & $^3S_1-^3D_1\,, \, U_{12}$ &  $^3P_2-^3D_2\,, \, U_{11}$\tabularnewline
\hline
$Q^0$&$-2.95\times 10^2  $&  &    \tabularnewline
\hline
$Q^1$&$6.14\times 10^2  $& &  \tabularnewline
\hline
$Q^2$&$2.08\times 10^2 $&$-5.92\times 10^3  $  &$5.59\times 10^2$ \tabularnewline
\hline
$Q^3$&$2.07\times 10^2 $&$-6.94 \times 10^3 $ &$7.50 \times 10^2$  \tabularnewline
\hline
\end{tabular}
\end{center}
\caption{Matrix elements of the generalized potential at the matching point $U_{ij}(\mu_M^2)$ that contain
adjusted parameters at different chiral orders after subtracting one-pion exchange contribution.
Coupled partial waves.}
\label{table:parameters2}
\end{table}

\bibliography{1}

\begin{thebibliography}{10}
\providecommand{\url}[1]{\texttt{#1}}
\providecommand{\urlprefix}{URL }
\providecommand{\eprint}[2][]{\url{#2}}

\bibitem{Weinberg:1991um}
S.~Weinberg, Nucl. Phys. \textbf{B363}, 3 (1991)

\bibitem{Ordonez:1992xp}
C.~Ordonez, U.~van Kolck, Phys.Lett. \textbf{B291}, 459 (1992)

\bibitem{Entem:2003ft}
D.~Entem, R.~Machleidt, Phys.Rev. \textbf{C68}, 041001 (2003),
  \eprint{nucl-th/0304018}

\bibitem{Epelbaum:2004fk}
E.~Epelbaum, W.~Gl{\"o}ckle, U.-G. Mei{\ss}ner, Nucl.Phys. \textbf{A747}, 362
  (2005), \eprint{nucl-th/0405048}

\bibitem{Epelbaum:2008ga}
E.~Epelbaum, H.-W. Hammer, U.-G. Mei{\ss}ner, Rev.Mod.Phys. \textbf{81}, 1773
  (2009), \eprint{0811.1338}

\bibitem{Machleidt:2011zz}
R.~Machleidt, D.~Entem, Phys.Rept. \textbf{503}, 1 (2011), \eprint{1105.2919}

\bibitem{Goldberger:1960md}
M.~Goldberger, M.~T. Grisaru, S.~MacDowell, et~al., Phys.Rev. \textbf{120},
  2250 (1960)

\bibitem{Ball:1965sa}
J.~S. Ball, A.~Scotti, D.~Y. Wong, Phys.Rev. \textbf{142}, 1000 (1966)

\bibitem{Scotti:1963zz}
A.~Scotti, D.~Wong, Phys.Rev.Lett. \textbf{10}, 142 (1963)

\bibitem{Scotti:1965zz}
A.~Scotti, D.~Wong, Phys.Rev. \textbf{138}, B145 (1965)

\bibitem{Albaladejo:2011bu}
M.~Albaladejo, J.~Oller, Phys.Rev. \textbf{C84}, 054009 (2011),
  \eprint{1107.3035}

\bibitem{Albaladejo:2012sa}
M.~Albaladejo, J.~Oller, Phys.Rev. \textbf{C86}, 034005 (2012),
  \eprint{1201.0443}

\bibitem{Gasparyan:2010xz}
A.~Gasparyan, M.~F.~M. Lutz, Nucl.Phys. \textbf{A848}, 126 (2010),
  \eprint{1003.3426}

\bibitem{Gasparyan:2010fb}
A.~Gasparyan, M.~Lutz, Fizika \textbf{B20}, 55 (2011), \eprint{1012.5948}

\bibitem{Gasparyan:2011yw}
A.~Gasparyan, M.~Lutz, B.~Pasquini, Nucl.Phys. \textbf{A866}, 79 (2011),
  \eprint{1102.3375}

\bibitem{Ditsche:2012fv}
C.~Ditsche, M.~Hoferichter, B.~Kubis, et~al., JHEP \textbf{1206}, 043 (2012),
  \eprint{1203.4758}

\bibitem{Danilkin:2011fz}
I.~Danilkin, L.~Gil, M.~Lutz, Phys.Lett. \textbf{B703}, 504 (2011),
  \eprint{1106.2230}

\bibitem{Danilkin:2012ua}
I.~Danilkin, M.~Lutz, S.~Leupold, et~al., Eur.Phys.J. \textbf{C73}, 2358
  (2013), \eprint{1211.1503}

\bibitem{Rentmeester:1999vw}
M.~Rentmeester, R.~Timmermans, J.~L. Friar, et~al., Phys.Rev.Lett. \textbf{82},
  4992 (1999), \eprint{nucl-th/9901054}

\bibitem{Birse:2003nz}
M.~C. Birse, J.~A. McGovern, Phys.Rev. \textbf{C70}, 054002 (2004),
  \eprint{nucl-th/0307050}

\bibitem{Rentmeester:2003mf}
M.~C.~M. Rentmeester, R.~G.~E. Timmermans, J.~J. de~Swart, Phys. Rev.
  \textbf{C67}, 044001 (2003), \eprint{nucl-th/0302080}

\bibitem{Birse:2007sx}
M.~C. Birse, Phys.Rev. \textbf{C76}, 034002 (2007), \eprint{0706.0984}

\bibitem{Shukla:2008sp}
D.~Shukla, D.~R. Phillips, E.~Mortenson, J.Phys. \textbf{G35}, 115009 (2008),
  \eprint{0803.4190}

\bibitem{Birse:2010jr}
M.~C. Birse, Eur.Phys.J. \textbf{A46}, 231 (2010), \eprint{1007.0540}

\bibitem{Kaplan:1996xu}
D.~B. Kaplan, M.~J. Savage, M.~B. Wise, Nucl.Phys. \textbf{B478}, 629 (1996),
  \eprint{nucl-th/9605002}

\bibitem{Lepage:1997cs}
G.~Lepage 135--180 (1997), \eprint{nucl-th/9706029}

\bibitem{Beane:2001bc}
S.~Beane, P.~F. Bedaque, M.~Savage, et~al., Nucl.Phys. \textbf{A700}, 377
  (2002), \eprint{nucl-th/0104030}

\bibitem{Nogga:2005hy}
A.~Nogga, R.~Timmermans, U.~van Kolck, Phys.Rev. \textbf{C72}, 054006 (2005),
  \eprint{nucl-th/0506005}

\bibitem{Birse:2005um}
M.~C. Birse, Phys.Rev. \textbf{C74}, 014003 (2006), \eprint{nucl-th/0507077}

\bibitem{PavonValderrama:2005wv}
M.~Pavon~Valderrama, E.~Ruiz~Arriola, Phys.Rev. \textbf{C74}, 054001 (2006),
  \eprint{nucl-th/0506047}

\bibitem{Epelbaum:2006pt}
E.~Epelbaum, U.-G. Mei{\ss}ner  (2006), \eprint{nucl-th/0609037}

\bibitem{Long:2011xw}
B.~Long, C.~Yang, Phys.Rev. \textbf{C85}, 034002 (2012), \eprint{1111.3993}

\bibitem{Danilkin:2010xd}
I.~Danilkin, A.~Gasparyan, M.~Lutz, Phys.Lett. \textbf{B697}, 147 (2011),
  \eprint{1009.5928}

\bibitem{Chew:1960iv}
G.~F. Chew, S.~Mandelstam, Phys. Rev. \textbf{119}, 467 (1960)

\bibitem{Stoica:2011cy}
S.~Stoica, M.~Lutz, O.~Scholten, Phys.Rev. \textbf{D84}, 125001 (2011),
  \eprint{1106.5619}

\bibitem{Stapp:1956mz}
H.~Stapp, T.~Ypsilantis, N.~Metropolis, Phys.Rev. \textbf{105}, 302 (1957)

\bibitem{Fettes:1998ud}
N.~Fettes, U.-G. Mei{\ss}ner, S.~Steininger, Nucl. Phys. \textbf{A640}, 199
  (1998), \eprint{hep-ph/9803266}

\bibitem{Fettes:2000gb}
N.~Fettes, U.-G. Mei{\ss}ner, M.~Mojzis, et~al., Annals Phys. \textbf{283}, 273
  (2000), \eprint{hep-ph/0001308}

\bibitem{Lutz:1999yr}
M.~Lutz, Nucl.Phys. \textbf{A677}, 241 (2000), \eprint{nucl-th/9906028}

\bibitem{Higa:2003jk}
R.~Higa, M.~Robilotta, Phys.Rev. \textbf{C68}, 024004 (2003),
  \eprint{nucl-th/0304025}

\bibitem{Passarino:1978jh}
G.~Passarino, M.~Veltman, Nucl.Phys. \textbf{B160}, 151 (1979)

\bibitem{Semke:2005sn}
A.~Semke, M.~Lutz, Nucl.Phys. \textbf{A778}, 153 (2006),
  \eprint{nucl-th/0511061}

\bibitem{Fuchs:2003qc}
T.~Fuchs, J.~Gegelia, G.~Japaridze, et~al., Phys.Rev. \textbf{D68}, 056005
  (2003), \eprint{hep-ph/0302117}

\bibitem{Kaiser:1997mw}
N.~Kaiser, R.~Brockmann, W.~Weise, Nucl.Phys. \textbf{A625}, 758 (1997),
  \eprint{nucl-th/9706045}

\bibitem{Friar:1999sj}
J.~L. Friar, Phys.Rev. \textbf{C60}, 034002 (1999), \eprint{nucl-th/9901082}

\bibitem{Timmermans:1990tz}
R.~G.~E. Timmermans, T.~A. Rijken, J.~J. de~Swart, Phys. Rev. Lett.
  \textbf{67}, 1074 (1991)

\bibitem{Baru:2010xn}
V.~Baru, C.~Hanhart, M.~Hoferichter, et~al., Phys.Lett. \textbf{B694}, 473
  (2011), \eprint{1003.4444}

\bibitem{Friar:2003yv}
J.~L. Friar, U.~van Kolck, G.~Payne, et~al., Phys.Rev. \textbf{C68}, 024003
  (2003), \eprint{nucl-th/0303058}

\bibitem{Krebs:2012yv}
H.~Krebs, A.~Gasparyan, E.~Epelbaum, Phys.Rev. \textbf{C85}, 054006 (2012),
  \eprint{1203.0067}

\bibitem{Epelbaum:2003gr}
E.~Epelbaum, W.~Gloeckle, U.-G. Meissner, Eur.Phys.J. \textbf{A19}, 125 (2004),
  \eprint{nucl-th/0304037}

\bibitem{Epelbaum:2003xx}
E.~Epelbaum, W.~Gloeckle, U.-G. Meissner, Eur.Phys.J. \textbf{A19}, 401 (2004),
  \eprint{nucl-th/0308010}

\bibitem{Bernard:1996gq}
V.~Bernard, N.~Kaiser, U.-G. Mei{\ss}ner, Nucl. Phys. \textbf{A615}, 483
  (1997), \eprint{hep-ph/9611253}

\bibitem{Stoks:1993tb}
V.~Stoks, R.~Kompl, M.~Rentmeester, et~al., Phys.Rev. \textbf{C48}, 792 (1993)

\bibitem{Girlanda:2010zz}
L.~Girlanda, M.~Viviani, Few Body Syst. \textbf{49}, 51 (2011)

\bibitem{SAID_NN}
{ N. A. Arndt, et al.}, SAID online program, http://gwdac.phys.gwu.edu

\bibitem{Epelbaum:2012ua}
E.~Epelbaum, J.~Gegelia, Phys.Lett. \textbf{B716}, 338 (2012),
  \eprint{1207.2420}

\end{thebibliography}
\bibliographystyle{epja}

\end{document}